\documentclass{jetpl}
\twocolumn
\usepackage{color}
\newcommand{\rev}[1]{{#1}}

\lat

\title{Hall conductivity as the topological invariant in phase space in the presence of interactions and non-uniform magnetic field }

\rtitle{Hall conductivity as the topological invariant in phase space in the presence of interactions ... }

\sodtitle{Hall conductivity as the topological invariant in phase space in the presence of interactions and non-uniform magnetic field}

\author{C.X. Zhang $^+$, M.\,A.\,Zubkov$^{+*}$\/\thanks{e-mail: zubkov@itep.ru, On leave of absence from Institute for Theoretical and Experimental Physics, B. Cheremushkinskaya 25, Moscow, 117259, Russia}}

\rauthor{C.X. Zhang, M.\,A.\,Zubkov }

\sodauthor{Zhang, Zubkov}

\address{$^+$ Physics Department, Ariel University, Ariel 40700, Israel}

\date{\today}

\abstract{The quantum Hall conductivity in the presence of constant magnetic field may be represented as the topological TKNN invariant. Recently the generalization of this expression has been proposed for the non - uniform magnetic field. \rev{The quantum Hall conductivity is represented as the topological invariant in phase space in terms  of the Wigner transformed two - point Green function.} This representation has been derived when the inter - electron interactions were neglected. It is natural to suppose, that in the presence of interactions the Hall conductivity is still given by the same expression, in which the non - interacting Green function is substituted by the complete two - point Green function \rev{ including the interaction contributions}. We prove this conjecture within the framework of the $2+1$ D tight - binding model of rather general type using the ordinary perturbation theory.  }


\begin{document}

\maketitle


\section{Introduction}
\label{Sect_Intro}
 Since the discovery of the Quantum Hall Effect (QHE), there were many attempts to understand the quantization of Hall conductivity $\sigma_H$.
The appearance of the universal integer values of the Hall plateaus prompts that  $\sigma_H$ has the topological meaning, i.e. it is related to a certain topological invariant, which is robust to the smooth modification of the system. Indeed, the seminal paper \cite{TKNN} shows that $\sigma_H$ may be expressed through the integral of Berry curvature over the occupied electronic states. This is the so - called TKNN (Thouless,  Kohmoto,  Nightingale, den Nijs) invariant \cite{Fradkin,Tong:2016kpv,Hatsugai,Hall3DTI}.  The corresponding expression is the topological invariant, i.e. it is not changed when the system is modified smoothly. However, it has been obtained for the constant magnetic fields only. Later it has been shown that in the absence of the inter - electron interactions the TKNN invariant for the intrinsic QHE (existing without external magnetic field) may be expressed through the momentum space Green's function \cite{Matsuyama:1986us,Volovik0} (see also Chapter 21.2.1 in \cite{Volovik2003}). Recently these two results have been extended to the case of magnetic field varying as a function of coordinates. The corresponding expression for $\sigma_H$ is the topological invariant in phase space expressed through the Wigner transformation of the two point Green function \cite{ZW2019}. The mentioned representations of $\sigma_H$ through the topological invariants were derived for the non-interacting systems. It is widely believed, that in the presence of interactions the intrinsic anomalous quantum Hall effect (AQHE) conductivity is given by the expression of    \cite{Matsuyama:1986us,Volovik0}, in which the non - interacting two - point Green function has been substituted by the two point Green function with the interaction corrections. In the $2+1$ D QED this has been proved in \cite{parity_anomaly,parity_anomaly_}. The corresponding property is now referred to as the non - renormalization of the parity anomaly in $2+1$ D Quantum Electrodynamics by the high orders of perturbation theory.
In the recent paper \cite{ZZ2019} we investigated the influence of interactions on the anomalous quantum Hall (AQHE) conductivity in the tight - binding models of the $2 + 1$ D topological insulator and the $3 + 1 D$ Weyl semimetal. 
the influence of interactions on the Hall conductivity in external magnetic field has been discussed widely in the past (see, for example \cite{Hall000,TKNN2,Altshuler0,Altshuler} and references therein), however, this consideration has been limited by the case of the constant magnetic field. In the present work, we investigate the influence of Coulomb interactions on the quantum Hall effect in the presence of the non-uniform magnetic field. We will prove to all orders of perturbation theory, that the corresponding Hall conductivity is given by the topological invariant (in phase space) of \cite{ZW2019}, in which the two - point (Wigner transformed) Green function is substituted by the complete one with the interactions taken into account. 


On the technical side we consider the tight - binding models with the Coulomb interactions between the electrons. We will use Wigner - Weyl formalism \cite{1,2,berezin,6} adapted in \cite{Z2016_1,FZ2019,SZ2018,ZK2017,KZ2018} to the lattice models of solid state physics combined with the ordinary perturbation theory.

\section{Hall current in varying magnetic field and varying electric potential}
\label{SectHall}
Let us discuss first the system with the interactions neglected.
For definiteness let us start from the Euclidean action of the $2+1$D tight-binding model of electrons under the action of varying magnetic field and varying electric potential, whose three -  potential is $A_{\mu}$
\begin{eqnarray}\label{action_3D}
&&S = \int d\tau \sum_{{\bf x,x'}}\bar{\psi}_{\bf x'}\Big(i(i \partial_{\tau} - A_3(i\tau,{\bf x}))\delta_{\bf x,x'}
- i{\cal D}_{\bf x,x'}\Big)\psi_{\bf x}   \label{S}
\end{eqnarray}
where
\begin{eqnarray}\label{lattice_difference}
{\cal D}_{\bf x,x'}&=&-\frac{i}{2} \sum_{i=1,2} [(1+\sigma^i)\delta_{x+e_i,x'}e^{iA_{x+e_i,x}}   \nonumber\\
&& + (1-\sigma^i)\delta_{x-e_i,x'}e^{iA_{x-e_i,x}}] \sigma_3    +i(m+2)\delta_{\bf x,x'}\sigma_3.  \nonumber
\end{eqnarray}
And we denote the parallel transporters along the lattice vector $e_i$ by $e^{iA_{x-e_i,x}}=e^{i\int_{x-e_i}^x A_\mu e_i^\mu du}$.

However, all results presented here are valid for any lattice models with the gauge invariant action of Eq. (\ref{S}), and Hermitian matrix $i{\cal D}_{\bf x, x'}$.
If vector potential $A_\mu(x)$ does not vary fast, i.e. if its variation on the distance of the lattice spacing may be neglected, then
Wigner transformation of the two-point Green function $G^{}_{W}(R,p)$ satisfies the Groenewold equation \cite{Z2016_1}
$$G^{}_{W}(R,p) * Q^{}_W(R,p)= 1$$
in which $$*=e^{\frac{i}{2}\overleftarrow{\partial}_x \overrightarrow{\partial}_p - \frac{i}{2} \overleftarrow{\partial}_p \overrightarrow{\partial}_x} $$ is the star (Moyal) product, where the derivatives with the left arrow act only on the functions standing to the left from the star while the derivatives with the right arrow act only to the functions standing right to the star, while for the model with the action of Eq. (\ref{S})
$$Q^{}_W(R,p)={\cal Q}(p-{ A}(R))$$
 $$= i\Big(\sum_{k=1,2,3}\sigma^k g_k(p-A(R))-im(p-A(R))\Big)\sigma^3$$ with $m(p) = m + \sum_{k=1,2}(1-{\rm cos}\,(p_k))$ and $g_k(p) = {\rm sin}\,p_k$ for $k=1,2$ while $g_3(p) = p_3=\omega$. Here $p = ({\bf p},p_3), R= ({\bf R},\tau)$. For the lattice model of a general type $\cal Q$ is a certain function specific for the given system. For our purposes it may be almost arbitrary.
Wigner transformation of the Green function $G(p_1,p_2)$ is defined as
$$
G_W(R,p) = \int_{\cal M} G(p + q/2,p-q/2)e^{iqx}dq
$$
where integral is over momentum space $\cal M$.  The star product is associative: $(f * g)*h = f*(g*h)$, which allows us to write such products without brackets.
The electric current density (in the absence of electric field) is given by
\begin {eqnarray}\label{current}
J^k(R) =\int \frac{d^3 p}{(2\pi)^3}  Tr G^{}_{W}(R,p) \frac{\partial}{\partial p_k} Q^{}_{W}(R,p).
\end{eqnarray}
For the convenience, we introduce the average total current $I_k= (T/S) \int J^k(R) d^3 R$, in which
$T$ is temperature, while $S$ is the area of the sample. In the following for simplicity  we refer to $I^k$ as to the total current. Under the periodic boundary conditions $I_k$ can be expressed as follows \cite{ZW2019}:
\begin {eqnarray}\label{current_ave}
I^k = \frac{T}{S}\int d^3 R \frac{d^3 p}{(2\pi)^3}  Tr G^{}_{W}(R,p)* \frac{\partial}{\partial p_k} Q^{}_{W}(R,p).
\end{eqnarray}
Here it is used that for any periodic functions in phase space $f$ and $g$, $\int dx dp f(x,p) * g(x,p) = \int dx dp f(x,p) g(x,p) $.
Let us consider the small modification of vector potential due to the extra (external) electric filed:
$A_{\mu} \rightarrow A_{\mu}+ \delta A_{\mu}$.
$\delta A_{\mu}$ corresponds to this extra small constant external electric field.
 We consider the variation of electric current $\delta I_k$ with respect to the variation $\delta A$, and obtain
\begin{eqnarray}\label{current_a}
I^k(A+\delta A) &=&T\int \frac{d^3 R}{S} \int \frac{d^3 p}{(2\pi)^3}  Tr G_{W}(R,p)  \nonumber\\&&\frac{\partial}{\partial p_k} {\cal Q}(p-{ A}(R)-\delta { A}),
\end{eqnarray}
where $G_{W}$ satisfies  $G_{W}(R,p) * Q_{W}(R,p)=1$, with
$Q_{W}(R,p)= {\cal Q}(p-{ A}(R)-\delta { A})$.
Contrary to Eq.(\ref{current_ave}), the Eq.(\ref{current_a}) does not contain the
star $*$, because the introduction of electric field breaks the periodical boundary conditions.
Using the expansion in powers of $\delta A$ we obtain ${\cal Q}(p-{A}(R)-\delta { A})\approx
{\cal Q}(p-{A}(R))-\partial^{\mu}{\cal Q}\delta {A}_{\mu}$, next we
 expand function $G_{W}(R,p)$ in powers of $\delta A$:   $G_{W}(R,p)=G^{(0)}_{W}+G^{(1)}_{W}+...$,
with $G^{(n)}_{W}\sim O([\delta A]^n)$.

In the leading (zeroth) order  $G^{(0)}_{W}$ satisfies equation
\begin {eqnarray}\label{WignerEqu_0}
G^{(0)}_{W}(R,p) * {\cal Q}(p-{ A}(R))= 1.
\end{eqnarray}
%
In the next (the first) order
$G^{(1)}_{W}$ satisfies
\begin {eqnarray}\label{WignerEqu_1}
0&=&G^{(1)}_{W}(R,p) * {\cal Q}(p-{A}(R))\nonumber\\&&-G^{(0)}_{W}(R,p)* \Big(\frac{\partial {\cal Q}(R,p)}{\partial p_{\mu}} \delta {A}^{\mu}\Big)
\end{eqnarray}
Solution of this equation gives
\begin {eqnarray}\label{WignerEqu_1}
G^{(1)}_{W}(R,p)&=&G^{(0)}_{W}(R,p)* \Big(\frac{\partial {\cal Q}(R,p)}{\partial p_{\mu}} \delta {\cal A}^{\mu}\Big)*G^{(0)}_{W}(R,p)    \nonumber\\
                        &=&(G^{(0)}_{W}* \partial_{\mu} {\cal Q}*G^{(0)}_{W}) \delta{\cal A}^{\mu}                                         \nonumber\\
                      &&+\frac{i}{2}(G^{(0)}_{W}* \partial_{\mu} {\cal Q}*\partial_{\nu}G^{(0)}_{W}) \partial_{x_{\nu}}\delta{\cal A}^{\mu}\nonumber\\&&
                          -\frac{i}{2}(\partial_{\nu}G^{(0)}_{W}* \partial_{\mu} {\cal Q}*G^{(0)}_{W}) \partial_{x_{\nu}}\delta{\cal A}^{\mu},
\end{eqnarray}
and the last line of the above equation may be transformed into
$(G^{(0)}_{W}* \partial_{\mu} {\cal Q}*\partial_{\nu}G^{(0)}_{W})
\delta F_{\mu\nu}$.
Therefore, we find the variation of current $\delta I_k $, up to the order of  $\delta A$ as follows
\begin {eqnarray}\label{current_b}
\delta I^k 
           &&=\frac{iT}{2}\delta F_{lm}  \int \frac{d^3 R}{S} \int \frac{d^3 p}{(2\pi)^3}Tr( \partial_l G^{(0)}_{W}\nonumber\\
           &&* \partial_m Q^{(0)}_{W}*G^{(0)}_{W})*
                                                                  \partial_k Q^{(0)}_{W},
\end{eqnarray}
Here we assume that the electric field strength $\delta F_{lm}$ is constant.

The last expression allows to obtain the following representation for the average Hall  conductivity (Electric field is directed along the $y$ axis):
$
\sigma_{xy} = \frac{\cal N}{2 \pi}
$,
where ${\cal N}$ is the topological invariant in phase space, which is the generalization of the  classical TKNN invariant \cite{TKNN}. Unlike the latter it is applicable to the non - homogeneous systems
\begin{eqnarray}
 {\cal N} & =&   \frac{T}{{ S}\,3!\,4\pi^2}\,  \epsilon_{ijk} \,
\int  d^3 x  \, \int d^3p \, Tr
  {G}_W(p,x)\ast \frac{\partial {Q}_W(p,x)}{\partial p_i} \nonumber\\&&\ast \frac{\partial  {G}_W( p,x)}{\partial p_j} \ast \frac{\partial  {Q}_W(p,x)}{\partial p_k}
	\label{calM2d23c}
\end{eqnarray}
This expression gives the average Hall conductivity in the presence of the non - homogeneous magnetic field and non - homogeneous electric potential, but with the interactions neglected. 

It is natural to suppose also, that Eq. (\ref{calM2d23c}) remains valid in the presence of the inter - electron interactions. Namely, one may suppose, that in the presence of interaction one simply has to substitute to Eq. (\ref{calM2d23c}) the complete two - point Green function with the contribution of interactions included. In the next sections we will prove this conjecture using the ordinary perturbation theory.

\section{$2+1$ D tight - binding model in the presence of Coulomb interactions. Setup of the {\it gedankenexperiment}.}

\label{SectCoulomb}


Below we consider the $2+1$ D tight-binding model with
Coulomb interactions. In order to apply the periodical boundary condition, we place our system into the surface of a large torus, or, equivalently, onto the cylinder closed through the spacial infinity. The coordinate system is attached to the surface of the cylinder:  the  x-direction is along its axis, and the y-direction is the circle with $y\in (-L,L]$. It is assumed that $L$ is much larger than any other physical parameter of the dimension of length existing in the given system. Therefore, we deal with the given system in a vicinity of any point as if the surface of the cylinder is flat.
We imagine that the cylinder is divided into the two parts:
(I) in the region $y\in [0,L]$ the effective fine structure constant $\alpha$ is nonzero, i.e. there are the Coulomb interactions between the electrons;
(II) in the region \rev{ $y\in (-L,0)$, } the effective fine structure constant $\alpha^\prime$ differs from that of the region (I). We will consider the limiting case $\alpha^\prime \to 0$, when there are no Coulomb interactions between the electrons.
We will consider the Hall current in this system in the presence of the non-uniform magnetic field, which is orthogonal to the surface of the cylinder. We will assume that the magnetic field varies around a constant value. In addition, it is supposed, that the profile of magnetic field in the piece (II) repeats its profile in the piece (I).  We may also assume the presence of varying electric potential. Then the same refers to the profile of electric potential. 

Vector potential $A_{\mu}$ is divided into the two contributions: $A_{\mu}=A^{m}_{\mu} + A^{e}_{\mu}$, where $A^{m}_{\mu}$ corresponds to the magnetic field and to the electric potential varying within the material, $A^{e}_{\mu}$ corresponds to external electric field, which is supposed to be small. The external electric field is uniform within each region, but it has opposite directions:
in the region (I), i.e. when $y\in [0,L]$, the electric field is along the positive y-direction, while in the region (II), where $y\in [-L,0]$,  the electric field is along the negative y-direction.
The Euclidean action is
\begin{eqnarray}
&&S = \int d\tau \sum_{{\bf x,x'}}\Big[\bar{\psi}_{\bf x'}\Big(i(i \partial_{\tau} - A_3(i\tau,{\bf x}))\delta_{\bf x,x'}
- i{\cal D}_{\bf x,x'}\Big)\psi_{\bf x}\nonumber\\
&&+\alpha \bar{\psi}(\tau,{\bf x})\psi(\tau, {\bf x})\theta(y)V({\bf x-x'})\theta(y')\bar{\psi}(\tau,{\bf x'})\psi(\tau, {\bf x'})\Big]
\end{eqnarray}
with the same function ${\cal D}_{\bf x,x'}$ as above and with $A_{x,y}=\int^y_x A^{\mu} ds_{\mu} $.
$V$ is the Coulomb potential $V({\bf x})=1/|{\bf x}|=1/\sqrt{x_1^2+x_2^2}$,
for ${\bf x}\not= {\bf 0}$.
Deep within the region (I) it might be more convenient to consider the action in momentum space, i.e.
%
$S= \int  dp \bar{\psi}_{p}\hat{Q}(p,i\partial_p)\psi_{p}
+\alpha \int  dp dq dk \bar{\psi}_{p+q}\psi_{p}\tilde{V}({\bf q})\bar{\psi}_{k}\psi_{q+k},$
%
where \cite{Z2016_1,SZ2018} $\hat{Q}(p,i\partial_p)=i\Big(\sum_{k=1,2,3}\sigma^k g_k(p-A(i\partial_p))-im(p-A(i\partial_p))\Big)\sigma^3$, and
%
$\tilde{V}({\bf q}) =\sum_{\bf x} \frac{e^{i{\bf q\cdot x}}}{\sqrt{x_1^2+x_2^2}}$.
%
The Coulomb interaction contributes to the self-energy of the fermions,
and the leading order contribution is proportional to $\alpha$.
The Green function can be calculated through the Feynman diagrams as follows
\begin {eqnarray}\label{Green_a}
&& G_{\alpha}(x,y) =  G_{0}(x,y) \nonumber\\&&+ \int G_{0}(x,z_1) \Sigma(z_1,z_2)G_{0}(z_2,y)dz_1 dz_2\nonumber\\&& + \int G_{0}(x,z_1) \Sigma(z_1,z_2)G_{0}(z_2,z_3)\nonumber\\&&\Sigma(z_3,z_4)G_{0}(z_4,y)dz_1 dz_2 dz_3 dz_4 +...
\end{eqnarray}
with
\begin {eqnarray}\label{Sigma_1}
\Sigma(z_1,z_2)&=&\alpha G_{0}(z_1,z_2) \theta(y_1)V({\bf z}_1-{\bf z}_2)\theta(y_2)+O(\alpha^2),\nonumber
\end{eqnarray}
where $\Sigma(z_1,z_2)=\alpha G_{0}(z_1,z_2) \theta(y_1)V({\bf z}_1-{\bf z}_2)\theta(y_2)+O(\alpha^2)$,
with $z_i=({\bf z}_i,\tau_i)$, and $z_i=(x_i,y_i)$.
After Wigner transformation, one finds that
\begin {eqnarray}\label{Green_b_Wigner}
&&G_{\alpha,W}(R,p) = G_{0,W}(R,p)\nonumber\\&& +  G_{0,W}(R,p)* \Sigma_W(R,p)* G_{0,W}(R,p) + ...,
\end{eqnarray}
where $G_{0,W}(R,p)$ satisfies $Q_{0,W}(R,p)* G_{0, W}(R,p) = 1$, equivalent to Eq.(4), while $\Sigma_W$ is Wigner transformation of $\Sigma$.

\section{Expression for the electric current through the interacting Green function}

\begin{figure}[h]
	\centering  %
	\includegraphics[width=5cm]{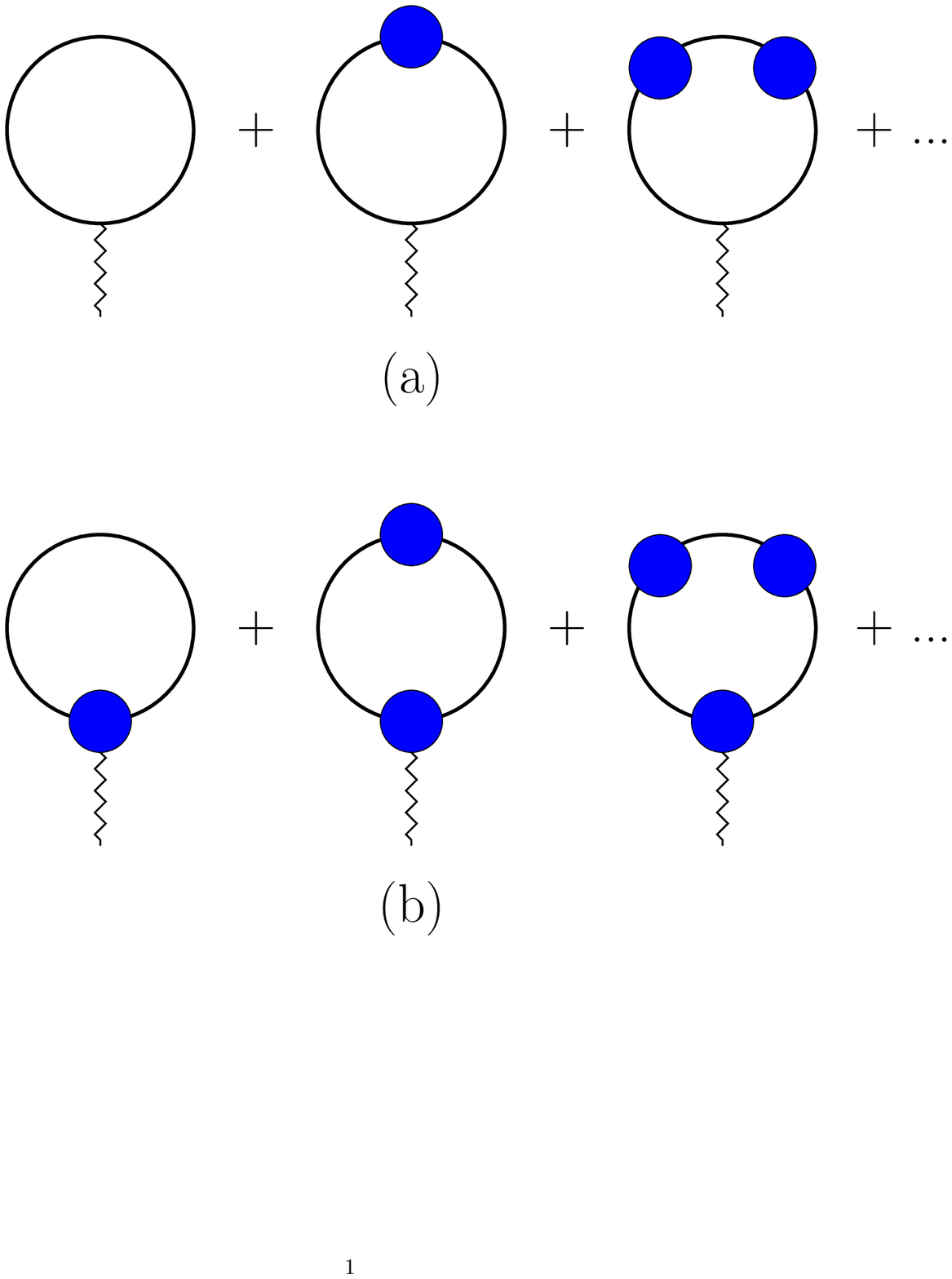}  %
	\caption{Figure 1. a) Feynmann diagrams  for $I^k(\alpha) = \int \frac{Td^3R}{S} \frac{d^3p}{(2\pi)^3} Tr G_{\alpha,W} \partial_{p_k} Q_{0,W}$ (expression for the electric current). The filled circles mark $\Sigma_W$. The external wavy line marks the position of $\partial_{p_k} Q_{0,W}$. b) Feynmann diagrams for $\Delta I^k(\alpha) = \int \frac{Td^3R}{S} \frac{d^3p}{(2\pi)^3} Tr G_{\alpha,W} \partial_{p_k} \Sigma_{W}$. The filled circle with the external wavy line marks $\partial_{p_k} \Sigma_{W}$.}  %
	\label{fig.3}   %
\end{figure}

Let us use the above developed technique for the calculation of the total electric current in the {\it Gedankenexperiment} under consideration.
It is convenient to expand $G_{\alpha,W}(R,p)$ in powers of the coupling constant $\alpha$
as $G_{\alpha,W}= {\cal G}_0 + \alpha {\cal G}_1 +\alpha^2 {\cal G}_2 +... $
Similarly, for each $G^{(l)}_{\alpha,W}$ in the series
$G_{\alpha,W}=  G^{(0)}_{\alpha,W}+  G^{(1)}_{\alpha,W}+ ...$
we have $G^{(l)}_{\alpha,W}=\sum_k \alpha^k {\cal G}^{(l)}_k $.
The total electric current may also be expanded in powers of $\alpha$ in both regions I and II and is given by
\begin {eqnarray}
I^k(\alpha) &=& \frac{T}{S}\int d^3 R \int \frac{d^3 p}{(2\pi)^3}  Tr G_{\alpha,W}(R,p) * \frac{\partial}{\partial p_k} Q_{0,W}(R,p)\nonumber\\
         &=&  \frac{T}{S} \int d^3 R \int \frac{d^3 p}{(2\pi)^3}  Tr (G_{0,W} \nonumber\\&&+ \sum_{n=1,2,...}G_{0,W}(*\Sigma_W*G_{0,W})^n)* \frac{\partial}{\partial p_k} Q_{0,W}(R,p)
         \label{current_3D_a}
\end{eqnarray}
We represent $\Sigma_W = \alpha \Sigma_{1,W} +\alpha^2 \Sigma_{2,W} +... $,
and the current is given by
 $I^{\mu}= I^{\mu}_0 + \alpha I^{\mu}_1 +\alpha^2 I^{\mu}_2 +... $,
in which $I^{k}_0= \frac{T}{S}\int d^3 R \int \frac{d^3 p}{(2\pi)^3}  Tr G_{0,W}*\frac{\partial}{\partial p_k} Q_{0,W}$,
%
and
\begin {eqnarray}\label{current_i}
I^k_r &=&\int \frac{Td^3 R d^3p}{S(2\pi)^3} Tr \sum_{k_1+...+k_n=r}\Big[\Pi_{i = 1...n} \Sigma_{k_i,W}*G_{0,W}\Big] \nonumber\\&&\frac{\partial}{\partial p_k} {\cal Q}_{0,W}G_{0,W},
\end{eqnarray}
with $r\geq 1$.
Let us compare the obtained expression for the total electric current with the following expression written through the interacting Green function
\begin{equation}
\tilde{I}^k(\alpha) = \int \frac{Td^3 R d^3p}{S(2\pi)^3}  Tr G_{\alpha,W}(R,p) * \frac{\partial}{\partial p_k} Q_{\alpha,W}(R,p)\label{tildeI}
\end{equation}
where $Q_{\alpha,W}(R,p)$ satisfies equation \begin{equation}
Q_{\alpha,W}(R,p) * G_{\alpha,W}(R,p) = 1 \label{groenewold}\end{equation}
For this purpose we calculate
$\Delta I^k(\alpha) = I^k(\alpha)-{\tilde I}^k(\alpha)$ 
 given by
\begin {eqnarray}
\Delta I^k &=& \int \frac{Td^3 R}{S} \int \frac{d^3 p}{(2\pi)^3}  Tr G_{\alpha,W}(R,p) * \frac{\partial}{\partial p_k} \Sigma_{W}(R,p)\nonumber\\
         &=& \int \frac{Td^3 R}{S} \int \frac{d^3 p}{(2\pi)^3}  Tr (G_{0,W} \nonumber\\&&+ \sum_{n=1,2,...}G_{0,W}(*\Sigma_W*G_{0,W})^n)* \frac{\partial}{\partial p_k} \Sigma_{\alpha,W}(R,p)\nonumber\\
         &=& \alpha \int \frac{Td^3 R}{S} \int \frac{d^3 p}{(2\pi)^3}  Tr G_{0,W}*\frac{\partial}{\partial p_k} \Sigma_{1,W}(R,p)\nonumber\\&&
            +\alpha^2 \int \frac{Td^3 R}{S} \int \frac{d^3 p}{(2\pi)^3} \Big(Tr \Sigma_{1,W} * G_{0,W}\nonumber\\&&*\frac{\partial}{\partial p_k}\Sigma_{1,W}(R,p) * G_{0,W} \nonumber\\&& +  Tr  G_{0,W}*\frac{\partial}{\partial p_k}\Sigma_{2,W}(R,p)\Big)+...\label{current_3D_a}
\end{eqnarray}
{The Feynmann diagrams corresponding to $\Delta I^k$ are represented in Fig. \ref{fig.3} b). Let us consider the diagram with $n$ self energies $\Sigma_W$:
\begin{eqnarray}
&&\Delta I^{(n)k} =  
 (n+1)\int \frac{Td^3 R d^3p}{S(2\pi)^3} Tr G_{0,W}*\partial_{p_k} Q_{0,W}* G_{0,W}...* \Sigma_{W} \nonumber\\
 &&-n\int \frac{Td^3 R d^3p}{S(2\pi)^3} Tr G_{0,W}*\partial_{p_k}\Sigma_W*...*\Sigma_{W}*G_{0,W}* \Sigma_{W}\nonumber
\end{eqnarray}
We come to the following relation
\begin{eqnarray}
&&(n+1)\Delta I^{(n)k} = (n+1)\int \frac{Td^3 R}{S} \frac{d^3p}{(2\pi)^3} Tr G_{0,W}*\partial_{p_k} Q_{0,W}\nonumber\\&&* G_{0,W}*...*\Sigma_{W}*G_{0,W}* \Sigma_{W},
\end{eqnarray}
which gives
$
\Delta I^{(n)k} = I^{(n+1)k}
$,
where $I^{(n+1)}$ is the contribution to electric current with $n+1$ insertions of $\Sigma_W$ represented schematically in Fig. \ref{fig.3} a) (the $n+2$ -th term in the sum). Overall, we obtain:
$$
\Delta I^k(\alpha) = I^k(\alpha)-I^{(0)k} = I^k(\alpha)-I^{k}(0)
$$
We find that the total current is given by an integral of Eq. (\ref{tildeI}) as long as the value of the total current remains equal to its value without interactions.  In the next section we will prove that indeed $I(\alpha) = I(0)$ in the region of analyticity in $\alpha$, i.e. as long as the perturbation theory in $\alpha$ may be used.}

\section{Non - renormalization of Hall conductance by interactions}

In the {\it Gedankenexperiment} under consideration the electric current in the absence of interactions is given by $I^k_0 =  \int \frac{Td^3 R}{S} \int \frac{d^3 p}{(2\pi)^3}  Tr G_{0,W}(R,p) * \frac{\partial}{\partial p_k} Q_{0,W}(R,p)$. Below we will prove that this expression does not receive corrections from interactions, i.e. for  $j \geq 1$, $I^k_j=0$.
\begin{figure}
			\includegraphics[height=3.5cm]{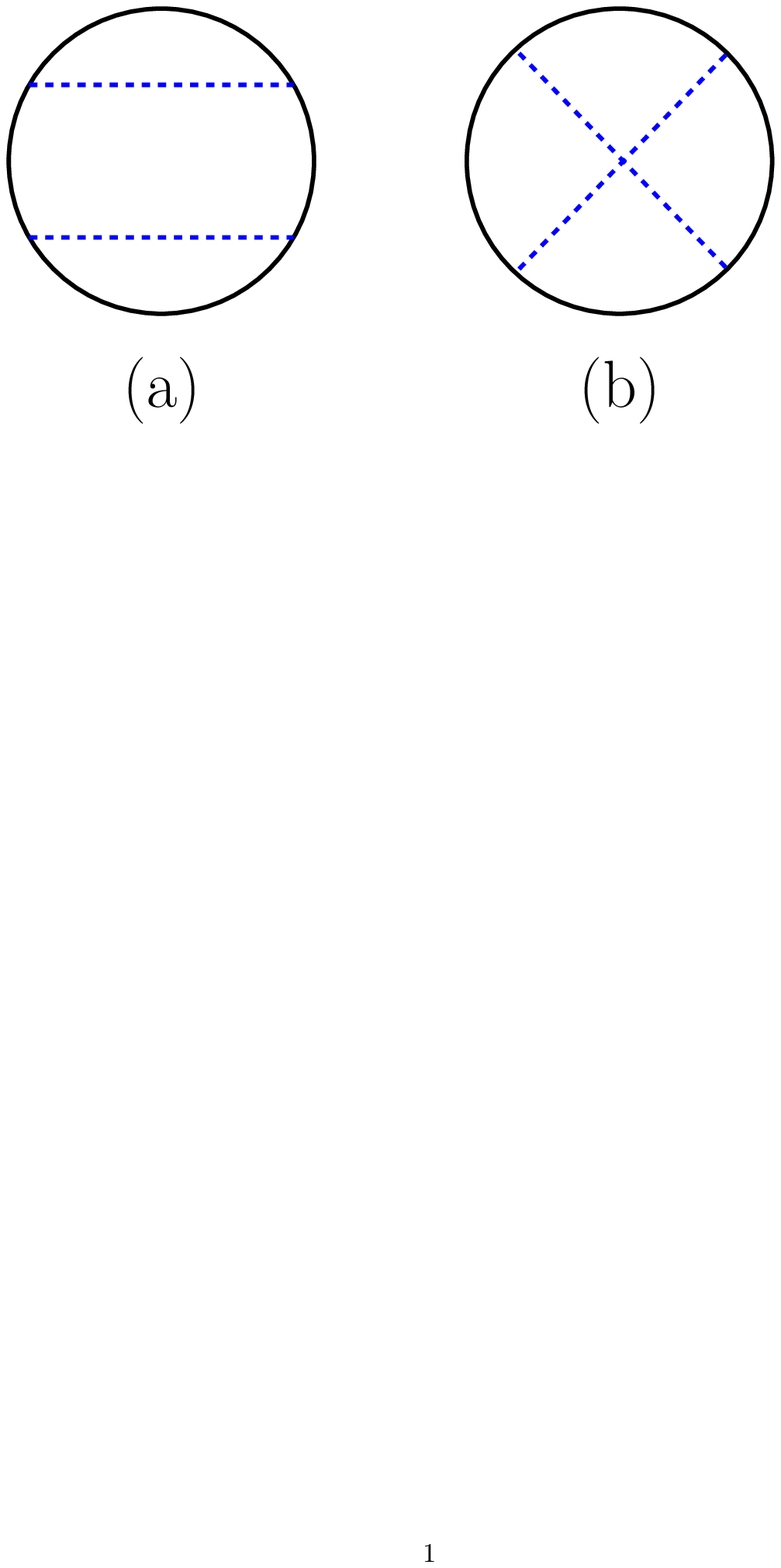}
	\caption{Figure 2. a) The progenitor diagram for the two - loop rainbow contribution to electric current. b) The progenitor diagram for the two - loop contribution to electric current (which is \rev{beyond the} rainbow approximation). 
}
	\label{progenitor}
\end{figure}
First, let us consider $I^k_1$,
which can be expressed explicitly as
\begin {eqnarray}\label{current_1st}
I^k_1 &=&-\int \frac{Td^3 R}{S} \int \frac{d^3 p d^3 q}{(2\pi)^6} Tr  ( G_{0,W}(R,p-q)D_W(R,q))
                                                               \nonumber\\&& * \frac{\partial}{\partial p_k} G_{0,W}(R,p) \nonumber\\
         &=&-\int \frac{Td^3 R}{S} \int \frac{d^3 p d^3 q}{(2\pi)^6} Tr  ( G_{0,W}(R,p-q)D_W(R,q))
                                                                \nonumber\\&&  \frac{\partial}{\partial p_k} G_{0,W}(R,p)         \end{eqnarray}
Here $D_W$ is the Wigner transformation of function
\begin {eqnarray}\label{Sigma_1}
D(z_1,z_2)&=&\alpha \theta(y_1)V({\bf z}_1-{\bf z}_2)\theta(y_2),\nonumber
\end{eqnarray}
We found that for each value of $R$ the above expression is proportional to
\begin {eqnarray}\label{lemma_1}
\int\int {\cal F}_R(p-q){\cal D}_R(q){\cal F}_R'(p)dpdq=0,
\end{eqnarray}
where ${\cal D}_R(q) = D_W(R,q)$ is an even function of $q$ while ${\cal F}_R(q) = G_{0,W}(R,q)$, and ${\cal F}'$ is the first derivative of $\cal F$. This representation allows us to prove that $I^k_1 =0$ (we perform the integration by parts and show that $I_1^k = -I^k_1$).
\begin{figure}[h]
	\centering  %
	\includegraphics[height=5cm]{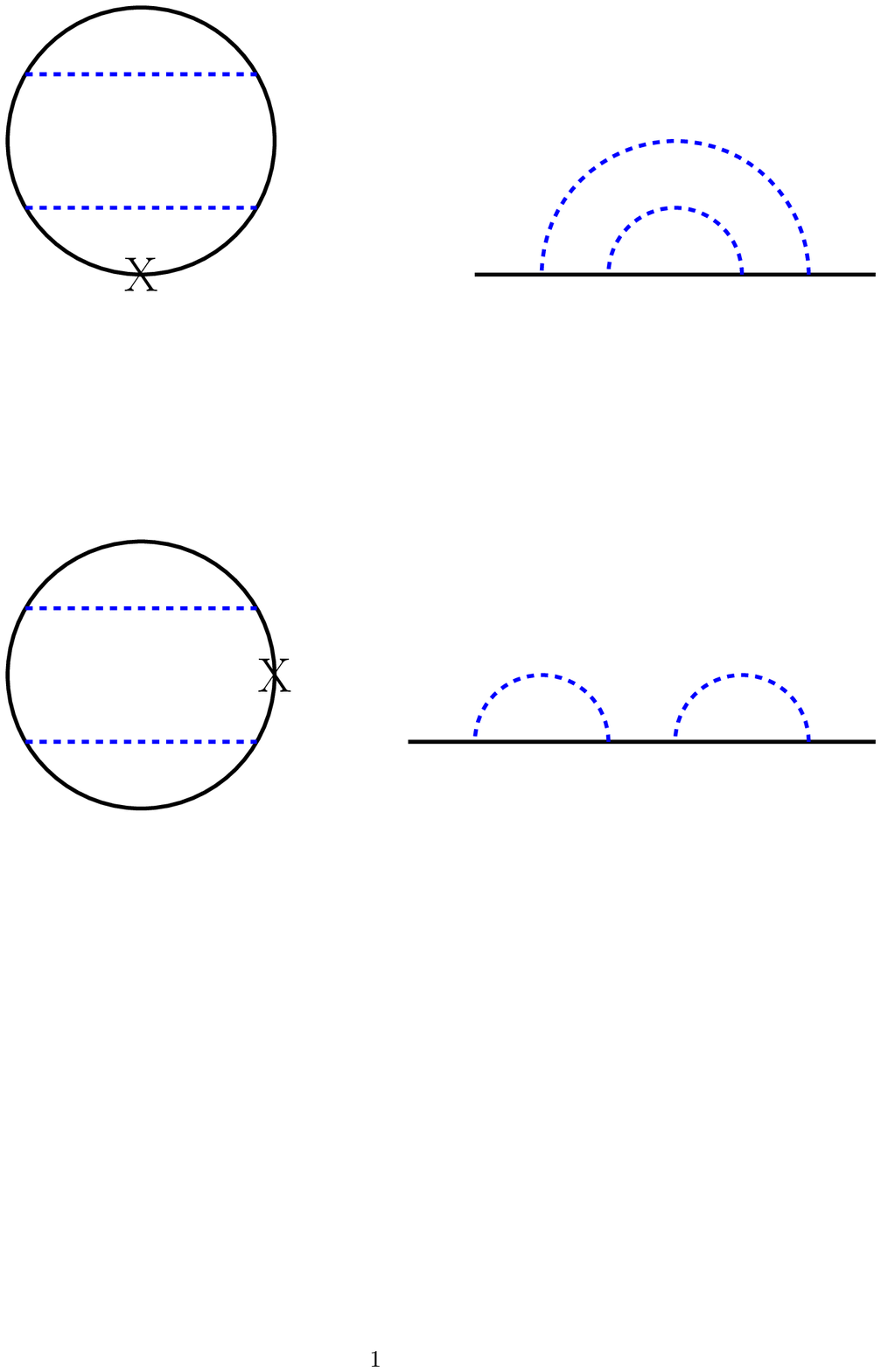} \vspace{1cm} %
	\caption{Figure 3. Two loop Feynmann diagrams for the self energy $\Sigma$ in rainbow approximation (right side of the figure) and the corresponding three loop rainbow contributions to electric current $I^k$ (left side of the figure). The crosses point out the positions of the derivatives $\partial_{p_k}Q_{0,W}$.}  %
	\label{fig.1}   %
\end{figure}
Let us now consider the next order contribution $I^k_2$. We have
\begin{eqnarray}\label{current_i2}
I^k_2 &=&-\int \frac{Td^3 R d^3p}{S(2\pi)^3} Tr \Sigma_{2,W} * \frac{\partial}{\partial p_k} G_{0,W}\nonumber \\&&
-\int \frac{Td^3 R d^3p}{S(2\pi)^3} Tr \Sigma_{1,W} * G_{0,W} * \Sigma_{1,W}* \frac{\partial}{\partial p_k} G_{0,W}\nonumber
\end{eqnarray}
Taking $\Sigma_2$ in rainbow approximation we get (see Fig. \ref{fig.1})
\begin{eqnarray}\label{current_i2}
 I^{k}_2 &\approx&-\int  \frac{T d^3 R d^3 pd^3 k d^3 q}{S(2\pi)^9} \,Tr \Big[G_{0,W}(R,p-k)\nonumber\\&&*G_{0,W}(R,p-k-q)D_W(R,q)* G_{0,W}(R,p-k)\Big]\nonumber\\&&D_W(R,k)*\partial_{p_k}G_{0,W}(R,p)  \nonumber\\&& -\int \frac{T d^3 Rd^3 pd^3 k d^3 q}{S(2\pi)^9}\,Tr G_{0,W}(R,p-q)D_W(R,q)\nonumber\\&&* G_{0,W}(R,p)*G_{0,W}(R,p-k)D_W(R,k)*\partial_{p_k}G_{0,W}(R,p)\nonumber
\end{eqnarray}
In the first term the star before $\partial_{p_k}$ may be eliminated. It may then be inserted before the last $D_W$ thus giving
\begin{eqnarray}\label{current_i2}
 I^{k}_2 &\approx&-\int \frac{T d^3 Rd^3 pd^3 k d^3 q}{S(2\pi)^9} \,Tr \Big[G_{0,W}(R,p-k)\nonumber\\&&*G_{0,W}(R,p-k-q)D_W(R,q)* G_{0,W}(R,p-k)\Big]*\nonumber\\&&D_W(R,k)\partial_{p_k}G_{0,W}(R,p)  \nonumber\\&& -\int \frac{T d^3 Rd^3 p}{S(2\pi)^3} \,Tr G_{0,W}(R,p-q)D_W(R,q)\nonumber\\&&* G_{0,W}(R,p)*G_{0,W}(R,p-k)D_W(R,k)\nonumber\\&&*\partial_{p_k}G_{0,W}(R,p)\nonumber\\&=&
 -\frac{1}{2}\int \frac{T d^3 Rd^3 pd^3 k d^3 q}{S(2\pi)^9}\,\partial_{p_k} \,Tr \Big[G_{0,W}(R,p-k)\nonumber\\&&*G_{0,W}(R,p-k-q)D_W(R,q)* G_{0,W}(R,p-k)\Big]*\nonumber\\&&D_W(R,k)G_{0,W}(R,p) = 0
\end{eqnarray}
Notice, that the last expression without derivative with respect to $p_k$ corresponds to the diagram similar somehow to the one called in \cite{parity_anomaly} "progenitor". We present the form of the corresponding Feynmann diagram in Fig. \ref{progenitor} a) and call it the progenitor for the diagrams presented in Fig. \ref{fig.1}. In essence, our present proof is an extension of the one given in \cite{parity_anomaly}.
The remaining two loop diagrams (see Fig. \ref{fig.2}) give the contribution that may be written as follows
\begin{eqnarray}\label{current_i2}
&& I^{k(cross)}_2 = -\int \frac{T d^3 Rd^3 pd^3 k d^3 q}{S(2\pi)^9} \,Tr \Big[G_{0,W}(R,p-k)\circ_{.2}\nonumber\\&&*G_{0,W}(R,p-k-q)\circ_{1.}D_{W(1)}(R,k)* G_{0,W}(R,p-q)\Big]\nonumber\\&&D_{W(2)}(R,q)\partial_{p_k}G_{0,W}(R,p) \nonumber\\
&&=-\frac{1}{4}\int \frac{T d^3 Rd^3 pd^3 k d^3 q}{S(2\pi)^9}\partial_{p_k} \,Tr \Big[G_{0,W}(R,p-k)\circ_{.2}\nonumber\\&&*G_{0,W}(R,p-k-q)\circ_{1.}D_{W(1)}(R,k)* G_{0,W}(R,p-q)\Big]\nonumber\\&&D_{W(2)}(R,q)*G_{0,W}(R,p) = 0
\end{eqnarray}
Here the star $*=e^{i\overleftarrow{\partial}_R\overrightarrow{\partial}_p/2-i\overleftarrow{\partial}_p\overrightarrow{\partial}_R/2}$ acts only on $G$ and does not act on $D$. \rev{ We denote by $\circ_{.i} = e^{i\overleftarrow{\partial}_R\overrightarrow{\partial}_p/2-i\overleftarrow{\partial}_p\overrightarrow{\partial}_R/2}$ the star product with derivatives over $p$ and $R$, in which the derivatives with the right arrow act on $D_{W(i)}$ while the derivatives with the left arrow act on the  fermion Green function standing to left from this symbol. Correspondingly, $\circ_{i.}$ is the star product, in which the derivatives with the right arrow act on the function standing immediately after this symbol while the derivatives with the left arrow act on $D_{W(i)}$. Notice, that since $D_{W(i)}$ does not contain $p$ the derivatives of $\circ$ act actually only on $D_{W(i)}$ and do not act on the corresponding $G$.}  The last line of the above expression corresponds to the diagram of Fig. \ref{progenitor} b).

One can see, that  $I_2^k=0$. In the same way the higher orders may be considered. 
One can check that $I^k_j=0$ for $j > 0$ to all orders of the perturbation theory.

\begin{figure}[h]
	\centering  %
	\includegraphics[height=2.5cm]{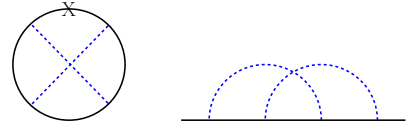}\vspace{1cm}  %
	\caption{Figure 4. Two loop Feynmann diagrams for the self energy $\Sigma$ \rev{beyond the} rainbow approximation (right side of the figure), and the corresponding three loop contributions to electric current $I^k$ (left side of the figure). The crosses point out the positions of the derivatives $\partial_{p_k}Q_{0,W}$.}  %
	\label{fig.2}   %
\end{figure}


The obtained results mean the following:
(1) The interaction corrections to the total electric current vanish in the system that contains the two pieces (with and without Coulomb interactions) in the presence of the electric field that is constant but has opposite directions in the two pieces of the material. We may think also, that in the first piece of the material the effective fine structure constant is nonzero while in the second piece it vanishes.
(2) There is the following representation for the total electric current in the considered system:
\begin {eqnarray}
&&{I}^k(\alpha) = \int \frac{Td^3 R d^3p}{S(2\pi)^3} Tr G_{\alpha,W}(R,p) * \frac{\partial}{\partial p_k} Q_{\alpha,W}(R,p)\nonumber\\&&=\int \frac{Td^3 R d^3p}{S(2\pi)^3}  Tr G_{\alpha,W}(R,p)  \frac{\partial}{\partial p_k} Q_{\alpha,W}(R,p)\label{IFIN}
\end{eqnarray}
The star is omitted in the last expression because the total system (containing  the two pieces) is defined with the periodical boundary conditions.

Now let us recall that in the piece of material without interactions the total electric current is proportional to electric field with the coefficient of proportionality (the conductivity) given by Eq. (\ref{calM2d23c}) divided by $2\pi$. As a result in the piece of the material with the interactions the total electric current is given by Eq. (\ref{IFIN}), in which the integral over $R$ is extended to the surface of the given piece only. The resulting expression leads to the Hall conductivity in this region:
$
\sigma_{xy} = \frac{\cal N}{2 \pi},
$
where ${\cal N}$ is the topological invariant in phase space given by
Eq. (\ref{calM2d23c}) with the complete Green function inserted instead of the noninteracting one. In turn, this expression for the conductivity appears to be equal to its value at $\alpha = 0$.

\section{Conclusions and discussion}

The results of the previous section demonstrate, that the (averaged over the system area) Hall conductivity in the presence of inhomogeneous magnetic field, inhomogeneous electric field, and Coulomb interactions is proportional to the topological invariant in phase space of Eq. (\ref{calM2d23c}). The present derivation of Eq. (\ref{calM2d23c}) (see also \cite{ZW2019,FZ2019} where this derivation has been given in the absence of interactions) is valid for the gauge field potential that varies slowly at the distances of the order of lattice spacing. This corresponds to the values of magnetic field much smaller than thousands Tesla and the wavelengths much larger than several Angstroms. In the region of analyticity in $\alpha$ the Hall conductivity does not depend on $\alpha$ at all and is still given by the same expression as without Coulomb interactions! To the best of our knowledge this result has been obtained for the first time for the systems in the presence of varying magnetic field. Previously the non - renormalization by interactions of the TKNN expression for $\sigma_H$ was proved for the case of the constant magnetic field only.



It is worth mentioning, that the problem considered here is technically more complicated than the case of the intrinsic Anomalous Quantum Hall effect (AQHE) existing in the $2D$ systems without magnetic field \cite{parity_anomaly,ZZ2019}. This is because in the latter case the fermion propagator (in the absence of disorder and external  electric field) depends on one conserved momentum, while in the presence of inhomogeneous magnetic field and electric potential we deal with the two - point Green function depending nontrivially on two momenta.    Our proof of the absence of radiative corrections to the Hall conductivity given above with slight modifications remains valid for the case of the AQHE as well. Moreover, it may be also  generalized to the other types of interactions (e.g. Yukawa, contact four-Fermi interaction, etc), and to the $3+1$ D systems as well. It would be interesting to consider the generalization of the approach of the present paper to the case, when elastic deformations are present (see, e.g. \cite{FZ2019}). In particular, in \cite{NV2018} it has been shown that the response of $\sigma_H$ to elastic deformations is quantized for the $3+1$D intrinsic AQHE in topological insulators. The influence of interactions on this response is worth to be considered.


The authors are grateful to I.Fialkovsky, M.Suleymanov, and Xi Wu for useful discussions. M.A.Zubkov kindly acknowledges valuable discussions with G.E.Volovik.%


\begin{thebibliography}{99}
\bibitem{TKNN} D. J. Thouless, M. Kohmoto, M. P. Nightingale, and M. den Nijs,
Phys. Rev. Lett. 49, 405 (1982).

\bibitem{Fradkin}
E. Fradkin, ``Field Theories of Condensed Matter Physics", 1991, Addison Wesley Publishing Company, Redwood City, CA

\bibitem{Tong:2016kpv} D. Tong, 
arXiv:1606.06687 [hep-ph]


\bibitem{Hatsugai}
Y. Hatsugai, 
J. Phys.: Condens. Matter 9 , 2507 (1997).


\bibitem{Hall3DTI}
X.-L. Qi, T. L. Hughes and S.-C. Zhang, Physical Review B 78, 195424 (2008).



\bibitem{Matsuyama:1986us}
  T.~Matsuyama,
  Prog.\ Theor.\ Phys.\  {\bf 77}  711 (1987).



\bibitem{Volovik0}
G.E. Volovik, 
JETP 67, 1804 (1988).

 \bibitem{Volovik2003}
G.E. Volovik, {\it The Universe in a Helium
Droplet}, Clarendon Press,  Oxford (2003).

\bibitem{ZW2019}
  M.~A.~Zubkov and X.~Wu,
  arXiv:1901.06661 [cond-mat.mes-hall].


\bibitem{parity_anomaly}
S. Coleman and B. Hill, Phys. Lett. B159 (1985) 184

\bibitem{parity_anomaly_}
T. Lee, Phys. Lett. B171 (1986) 247

\bibitem{ZZ2019}
  C.~X.~Zhang and M.~A.~Zubkov,
  arXiv:1902.06545 [cond-mat.mes-hall].


























\bibitem{Hall000}
Ryogo Kubo, Hiroshi Hasegawa, Natsuki Hashitsume,
Journal of the Physical Society of Japan 14(1) (1959) 56-74 DOI:
10.1143/JPSJ.14.56

\bibitem{TKNN2}
Q. Niu, D. J. Thouless,  and Y. Wu,
Phys. Rev. B 31, 3372  (1985).






\bibitem{Altshuler0}
B. L. Altshuler,
D. Khmel'nitzkii, A. I. Larkin and
P. A. Lee, 
Phys.Rev.B 22, 5142  (1980).


\bibitem{Altshuler}
B.L. Altshuler and A.G. Aronov,
Electron-electron inter-action in disordered systems (Editors: A.L. Efros, M. Pollak, Amsterdam, 1985).











 \bibitem{1}
 H. J. Groenewold, 
 Physica, 12, 405 (1946) .

 \bibitem{2}
 J. E. Moyal, 
 Proceedings of the Cambridge Philosophical Society, 45, 99 (1949).


 \bibitem{berezin}
F.A. Berezin  and M.A. Shubin, in: Colloquia Mathematica Societatis Janos Bolyai (North-Holland, Amsterdam) p. 21, (1972).


 \bibitem{6}
T. L. Curtright and C. K. Zachos,  
 Asia Pacific Physics Newsletter, issue 01, pages  37 $-$ 46 (2012),
arXiv:1104.5269.

\bibitem{Z2016_1}
  M.~A.~Zubkov,
  Annals Phys.\  {\bf 373}, 298 (2016).
  [arXiv:1603.03665 [cond-mat.mes-hall]].


\bibitem{FZ2019}
I. V. Fialkovsky, M. A. Zubkov,
arXiv:1905.11097

\bibitem{SZ2018}
  M.~Suleymanov and M.~A.~Zubkov,
  Nucl.\ Phys.\ B {\bf 938} , 171 (2019) (Corrigendum: https://doi.org/10.1016/j.nuclphysb.2019.114674 )
  [arXiv:1811.08233 [hep-lat]].














\bibitem{ZK2017}   M.~A.~Zubkov and Z.~V.~Khaidukov,  
      JETP Lett.\  {\bf 106},  172 (2017)    [Pisma Zh.\ Eksp.\ Teor.\ Fiz.\  {\bf 106} (2017) no.3,  166].

\bibitem{KZ2018}
  Z.~V.~Khaidukov and M.~A.~Zubkov,
  JETP Lett.\  {\bf 108} (2018) no.10,  670
  doi:10.1134/S0021364018220046
  [arXiv:1812.00970 [cond-mat.mes-hall]].






\bibitem{NV2018}
J. Nissinen and G.E.
Volovik, arXiv:1812.03175


\end{thebibliography}
\end{document}